\documentclass[aps,prc,onecolumn,superscriptaddress,showkeys]{revtex4-1}
\usepackage{amsmath,amssymb,bm}
\usepackage{graphicx}
\usepackage{multirow}
\usepackage{color}
\usepackage[normalem]{ulem}

\newcommand{\eq}[1]{\begin{equation} #1 \end{equation}}

\newcommand{\vect}[1]{\mathbf{#1}}
\newcommand{\vectop}[1]{\hat{\bm{#1}}}

\makeatletter
\def\p@subsection{}
\makeatother

\begin{document}

%\preprint{}

\title{Consistency of two different approaches to determine the
	strength of a pairing residual interaction in the rare-earth region}


%\author{Koh Meng-Hock}
%\thanks{}

\author{Nurhafiza M. Nor}
\affiliation{Department of Physics, Faculty of Science, Universiti
  Teknologi Malaysia, 81310 Johor Bahru, Johor, Malaysia}

\author{Nor-Anita Rezle}
\affiliation{Department of Physics, Faculty of Science, Universiti Teknologi Malaysia, 81310 Johor Bahru, Johor, Malaysia}

\author{Kai-Wen Kelvin-Lee}
\affiliation{Department of Physics, Faculty of Science, Universiti Teknologi Malaysia, 81310 Johor Bahru, Johor, Malaysia}

\author{Meng-Hock Koh}
\affiliation{Department of Physics, Faculty of Science, Universiti Teknologi Malaysia, 81310 Johor Bahru, Johor, Malaysia}
\affiliation{UTM Centre for Industrial and Applied Mathematics, 81310 Johor Bahru, Johor, Malaysia}

\author{L. Bonneau}
\affiliation{Universit\'{e} de Bordeaux, CENBG, UMR5797, F-33170 Gradignan, France}
\affiliation{CNRS, IN2P3, CENBG, UMR5797, F-33170 Gradignan, France}
%\email{bonneau@cenbg.in2p3.fr}

\author{P. Quentin}
\thanks{Corresponding author: philippe.quentin@tdtu.edu.vn}
\affiliation{Division of Nuclear Physics, Advanced Institute of Materials Science, Ton Duc Thang University, Ho Chi Minh City, Vietnam}
\affiliation{Faculty of Applied Sciences, Ton Duc Thang University, Ho Chi Minh City, Vietnam}
\affiliation{Universit\'{e} de Bordeaux, CENBG, UMR5797, F-33170 Gradignan, France}

\date{\today}

\begin{abstract}
Two fits of the pairing residual interaction in the rare-earth region
are independently performed. One is made on the odd-even staggering
of masses by comparing measured and explicitly calculated
three-point binding-energy differences centered on odd-even
nuclei. Another deals with the moments of inertia of the first
$2^{+}$ states of well deformed even-even nuclei upon comparing
experimental data with the results of Inglis-Belyaev moments
(supplemented by a crude estimate of the so-called Thouless-Valatin
corrections). The sample includes 24 even-even and 31 odd-mass
nuclei selected according to two criteria: they should have
good rotor properties and should not correspond to low 
pairing-correlation regimes in their ground states. Calculations are performed
in the self-consistent Hartree-Fock plus BCS framework (implementing a
self-consistent blocking in the case of odd-mass nuclei). The Skyrme SIII
parametrization is used in the particle-hole channel and the fitted
quantities are the strengths of $|T_z| = 1 $ proton and neutron seniority
residual interactions. As a result the two fits yield sets of strengths in
excellent agreement: about $0.1 \%$ for the neutron parameters and $0.2\%$ for protons. In contrast when one performs such a fit on
odd-even staggering from quantities deduced from BCS gaps or 
minimal quasiparticle energies in even-even nuclei, as is traditional,
one obtains results significantly different from those obtained in
the same nuclei by a fit of moments of inertia. As a conclusion,
beyond providing a phenomenological tool for microscopic calculations
in this region, we have illustrated the proposition made in the
seminal paper of Bohr, Mottelson and Pines that moments of inertia and
odd-even staggering in selected nuclei were excellent measuring sticks of
nuclear pairing correlations. Furthermore we have assessed the
validity of our theoretical approach which includes simple yet
apparently reasonable assumptions (seniority residual interaction,
parametrization of its matrix elements as functions of the nucleon
numbers and global Thouless-Valatin renormalisation of Inglis-Belyaev
moments of inertia).
\end{abstract}
%\keywords{}

\maketitle


\section{Introduction \label{Introduction}}
Any phenomenological approach of a property of physical interest
relies on a safe fitting process of the parameters of the theory which
attempts to describe it. To do so, two necessary conditions are
required: i) one must choose a quantity to be reproduced which is, to
a large extent, solely dependent on the property under study, ii) this
quantity should vary with respect to the fitting parameters in a fast
monotonic fashion. More precisely the range of the fitted parameters
corresponding to the relevant experimental error bars should be
considered as being small from the point of view of some other
physical considerations.

In this study we want to describe the spectroscopic properties of
rare-earth nuclei within a self-consistent BCS-type 
approach. Taking for granted that we have at our disposal a good
effective nucleon-nucleon interaction in the particle-hole channel, an
important challenge is thus to fit the parameters of the pairing
residual interaction. In the rare-earth region, the nuclei are 
far enough from the $N = Z$ line so that, as well known and easily checked,
neutron-proton pairing is inoperative. We will therefore restrict our study 
to the consideration of neutron-neutron, proton-proton (or $\vert T_z
\vert = 1$) pairing residual interactions.

This fit will be achieved in two independent ways: reproducing either the
odd-even staggering (OES) in ground-state energies, or the moment of
inertia of well and rigidly deformed nuclei. It is remarkable that
these two properties have been singled out as good indices of pair
correlations in the seminal paper of Bohr, Mottelson and Pines
\cite{Bohr_1958} on the existence of pair correlated nuclear
states analogous
to superconducting metallic states. These properties are quoted there after the
first evidence which is presented, namely the difference 
of particle-excitation nuclear spectra between even-even and 
odd-mass systems. While these differences in nuclear spectra
are rather difficult to reproduce theoretically in a systematic
fashion, the OES and moments of inertia are now within reach in
tractable and reliable calculations and thus well-adapted to a fitting
process.

We will demonstrate that the two approaches lead to consistent
results, thus 
substantiating at the same time the theoretical underlying
assumptions and their modelisation. 

\section{Principles of the fits \label{Principles}}

\subsection{Odd-even staggering of binding energies}

Traditionnally, it has been considered that a theoretical description
of pairing-correlation properties should be adjusted in such a way as
to reproduce the OES observed in ground-state energies. This energy
staggering has been  associated approximately with the BCS gap
parameter corresponding to the single-particle state of the unpaired nucleon
already, as we saw, from the beginning \cite{Bohr_1958} and this is
regularly quoted as such in textbooks (see, e.g., \cite{Bohr_1998,
  Ring_1980}). Fitting the pairing residual interaction parameters has
thus consisted in an attempt to reproduce as best as possible in a BCS
framework pairing gaps deduced through some finite difference formulae
(see below the discussion on how this is achieved) from the
consideration of the ground-state binding-energy surface $E(N,Z)$ of
nuclei with $N$ neutrons and $Z$ protons (see, e.g.,
Ref.~\cite{Moller_1992}). This fitting protocol has been followed also
in extensive self-consistent Hartree-Fock plus BCS calculations from
their beginning (see Ref.~\cite{Beiner_1970}) and on in many instances
as quoted for example in the rewiew paper of Ref.~\cite{Bender_2003}.

Within the BCS framework, we must be more specific. The simplest approach deals with constant pairing matrix elements
of the so-called seniority residual interaction
\begin{equation}\label{101}
g^{(q)} = \langle i \overline{i} \vert \widehat{v}_{\rm res} \vert j \overline{j} \rangle
- \langle i \overline{i} \vert \widehat{v}_{\rm res} \vert  \overline{j} j \rangle
\end{equation}
where the labels $i$ and $j$ refer to canonical basis states of the
charge state $q$ and $\widehat{v}_{\rm res}$ is the
residual interaction operator (as defined, e.g., in Ref.~\cite{LeBloas_2012}). 
Indeed one neglects in that case the state dependence of
these matrix elements, with the necessity of an energy cut-off of the
otherwise divergent corresponding calculations. As a consequence, this
cut-off is a primary parameter of the theory. Once this parameter is fixed, one
fits $g^{(q)}$ by equating the corresponding pairing gap
$\Delta^{(q)}$ (identical for each canonical basis state of charge
$q$) with some version of the OES energy. Alternatively, one may fit
(see, e.g., Ref. ~\cite{Bonche_1985}) this OES  energy with the
minimal quasi-particle (qp) energy in which the single-particle
(sp) energy is noted as $e_i$ 
\begin{equation}\label{102}
E_{\rm qp}^{(q)}(i) = \sqrt {(e_i - \lambda^{(q)})^2 + (\Delta^{(q)})^2} 
\end{equation}
where $\lambda^{(q)}$ is the corresponding chemical potential. 
One introduces thus a somewhat uncontrollable term $(e_i -
\lambda^{(q)})^2$.

A more advanced approach uses a spin-singlet zero-range (delta)
local interaction
\begin{equation}\label{103}
\hat v_\delta \propto 
\frac{1}{4} \, (1 - \vectop{\sigma}_1 \cdot 
\vectop{\sigma}_2) \, \delta(\vect r_1 - \vect r_2) 
\end{equation}
where $\vectop{\sigma}_i$ are spin Pauli matrices.
In line with the richer structural
properties of this interaction, its use in a BCS formalism induces a
state dependence of the pairing gaps. As a consequence the question of
knowing which sp configuration is to be chosen for the unpaired
particle becomes an important issue. Generally, one chooses the one
yielding the lowest qp energy, yet sometimes at the price of
describing an intrinsic configuration which might be different from
the experimental one. 

These calculations have been generally performed, at least until
rather recently, for even-even nuclei. 
This entails a priori two
deficiencies. Whatever the exact definition of the 
OES energy, one obviously 
has to deal 
with odd-neutron or odd-proton nuclei. 
In these systems the pairing is quenched
by the Pauli reduction of available levels onto which the residual
interaction can perfom pair transfers. 
Consequently, the pairing
correlations in even-even nuclei, and thus the corresponding gaps, are
overestimated with respect 
to what they are in the adjacent odd systems. The second drawback is 
related to the mean-field effect
affecting the energy differences between two neighbouring nuclei. 
Indeed the mean field can influence pairing
properties by changing the sp level
density at the Fermi surface  first by polarisation 
effect. This may lead
to different equilibrium deformations. Moreover,  
the mean field may
affect the sp level density 
as a consequence of the slight breaking of the
time-reversal symmetry  
resulting from an odd number
of fermions. In systems with such a number of nucleons 
the self-consistency of the mean-field removes the Kramers degeneracy
of conjugate single-particle states as discussed, e.g., in 
Refs.~\cite{Pototzky_2010,Koh_2016,Koh_2017}.

To minimize the polarisation effect, one must not rely on 
OES experimental estimates involving too long
isotopic or isotonic series, since, particularly in transitional
regions, they may involve too large variations of sp level
densities. One will thus preferably fit a three-point mass difference
formula. As discussed in Refs.~\cite{Dobaczewski_2001,
  Duguet_2001} such differences $\Delta^{(3)}_q$ 
centered around an odd-neutron (odd-proton resp.) are indeed good
markers of the neutron (proton resp.) degree of pairing
correlations. They are, to a large extent, free from 
single-particle
filling effects. Indeed, they are given for
  instance for an isotopic series by
\begin{subequations}
\begin{align}
\label{104} 
\Delta_n^{(3)}(N) & = \frac{(-1)^N}{2} \, \Big[E(N+1,Z)-2 \, E(N,Z) +
E(N-1,Z)\Big] \\
& =\frac{(-1)^N}{2} \Big[S_n(N,Z) - S_n(N+1,Z)\Big]
\end{align}
\end{subequations}
where $N$ is odd and $S_n(N,Z)$ is the experimental neutron separation energy of
a nucleus composed of $N$ neutrons and $Z$ protons.  

From the above one sees that centering the binding-energy 
difference on an odd-$N$ value prevents 
from unwanted energy jumps in the separation-energy
  differences caused by the occupation of different sp states
for the ejected nucleon.

In an approach where the fit is made on energy gaps (or qp energies),
however, one does not evaluate directly observable quantities. In this
paper, we compute explicitly OES energies, namely $\Delta^{(3)}_q$ differences. 
This implies computing
total ground-state energies of three adjacent nuclei 
(either isotopes or isotones), specifically two 
even-even nuclei and one odd-mass nucleus. We perform
  these calculations within the
Hartree-Fock plus BCS framework 
with self-consistent blocking for odd-mass nuclei. In this 
approach we take into account time-odd components in the mean-field, when needed. 
Even though, as above discussed, the
$\Delta^{(3)}_q$ terms are mostly dependent on pairing properties, we
can incorporate in such a way small possible polarisation effects.

\subsection{Moments of inertia of well and rigidly deformed nuclei}

As noted in Ref.~\cite{Bohr_1958} the quenching of the moments of
inertia of well and rigidly deformed even-even nuclei from their
rigid-body values constitute a clear manifestation of the existence of
pair correlations. It has received a physical explanation in terms of
a gradual alignment of the members of the Cooper pairs, dubbed as the Coriolis
anti-pairing effect in Ref.~\cite{Mottelson_1960}. This effect has
been introduced phenomenologically to modify the Inglis formula
\cite{Inglis_1954} in Ref.~\cite{Griffin_1960} by inserting a pairing
gap in the energy denominator. It has found later a sound theoretical
basis within the context of a microscopic Routhian approach \`a la
Thouless-Valatin~\cite{Thouless_1962}, by Belyaev~\cite{Belyaev_1961}
for rotations in an adiabatic regime. The resulting so-called
Inglis-Belyaev formula for the moments of inertia corresponds however
to a non-selfconsistent approximation of the adiabatic Time-Dependent
Hartree-Fock-Bogolyubov (ATDHF) approach of Baranger and V\'en\'eroni
\cite{Baranger_1978}. As discussed in Ref.~\cite{Yuldashbaeva_1999}
and more recently in Ref.~\cite{Baran_2011}, it does not take into
account the time-odd mean-field part brought in by the time-odd
component of the density matrix generated by the collective motion. It
has been shown~\cite{Yuldashbaeva_1999} that this omission entails a
spurious reduction of the ATDHF moment of inertia estimated on average
in Ref.~\cite{Libert_1999} to be approximately equal to 32\%. This
enhancement of the Inglis-Belyaev moments will be referred to as the
Thouless-Valatin correction.

As it has been clear from the first extensive calculations within the
Inglis-Belyaev framework (see, e.g., Ref.~\cite{Prior_1968}) the moments
of inertia are strongly dependent on the pairing
correlations. Increasing these
  correlations lead to a fast decrease of these
moments through the correlation-generated counter-rotating intrinsic
currents. Therefore moments of
  inertia qualify for the fit considered in this paper.

Specifically we will fit the moment of inertia of the first 2$^{+}$
states of well and rigidly deformed even-even nuclei in the
rare-earth region. 
The choice of such nuclear states is of course prompted by the
necessity to compare the above calculated adiabatic inertia parameters
with the nuclear states having the lowest available non-vanishing
angular velocity. In order for this comparison to make sense, one
should also make sure that the energy of this 2$^{+}$ state
corresponds to a pure rotational excitation mode. This implies that
the quantal shape fluctuations around the classical equilibrium
deformation are limited so that the description of this nuclear state
by a single BCS wavefunction makes some sense. 
%\sout{It has recently been
%quantitatively shown from microscopically based Bohr hamiltonian
%calculations~\cite{Rebhaoui_2018} that the many isotopes in this
%region which are well deformed (having intrinsic charge $\hat{Q}_{20}$
%expectation values of 7 barns or more) and which may be considered as
%good rotors (having a ratio $E_{4/2}$ of the energies of the first
%4$^{+}$ and 2$^{+}$ states in the 3.3 range) do not show indeed any
%significant coupling of the rotational and $\beta$ or $\gamma$
%vibrational modes in their first 2$^{+}$ states.} 
To assess this approximation, microscopically-based
  Bohr hamiltonian calculations of low-energy spectra have been
  recently performed in this mass region by Rebhaoui and
  collaborators~\cite{Rebhaoui_2018}. Many rare-earth isotopes are
  indeed well deformed, having intrinsic charge expectation values
  $\hat{Q}_{20}$ of 7 barns or more, and may be considered as good
  rotors, with a ratio $E_{4/2}$ of the energies of the first 4$^{+}$
  and 2$^{+}$ states in the 3.3 range. Rebhaoui and collaborators
  showed that these isotopes do not show any significant coupling of
  the rotational modes with $\beta$ or $\gamma$ vibrational modes in their
  first 2$^{+}$ states. As a conclusion, 
  the moment of inertia, being strongly dependent on
  pairing correlations, satisfies the two 
  criteria for a good fitting process mentioned at the
  beginning of the introduction. 

\section{Theoretical approach 
	\label{Theoretical approach}}

Our theoretical approach is based on the self--consistent Hartree--Fock--BCS framework yielding
an intrinsic state solution for the nuclei of interest. A phenomenological
Skyrme effective  nucleon-nucleon interaction is used. Axial and
intrinsic parity symmetries are assumed.

Calculations of even-even nuclei are performed according to the
standard method described in Ref.~\cite{Flocard_1973}, while in the
case of odd-mass nuclei, two approaches may be considered.

One is dubbed as the self-consistent blocking (SCB) framework. Within
this framework 
the single-particle state occupied by the unpaired nucleon is
blocked by setting its occupation probability to 1 while the occupation of its quasi-pair partner (as
defined below) is set to 0. These single-particle
states do not participate in the BCS pair-transfer process. The
time-reversal symmetry breaking inherent to the description of a
system with an odd number of fermions is reflected in the Hartree--Fock
field by the presence of time-odd terms which are defined within the
Skyrme formalism in terms of time-odd densities such as current and
spin-vector densities among others (see, e.g.,
Ref.~\cite{Hellemans_2012} for details). The two quantum numbers $K$
and $\pi$, respectively projection of the total angular momentum on
the symmetry $z$-axis and parity, are taken as those of the
experimental $I^{\pi}$ quantum numbers of the nuclear state which we
want to describe. The assimilation of the $K$ quantum number to the
total spin $I$ is made here upon assuming the validity of the
Bohr-Mottelson Unified Model description of rotational band heads in
deformed nuclei in the absence of Coriolis coupling.
 
Our restricted Bogoliubov qp 
transformation implies quasi-pairs consisting in couples of almost 
time-reversed states. These pairs are defined without ambiguity as 
described, e.g., in Refs.~\cite{Pototzky_2010,Koh_2016} due to the
small character of  the time-reversal symmetry breaking resulting from
the odd number of nucleons in such heavy nuclei.

In the second approach, called the \textit{equal filling
  approximation} (EFA), one sets the occupation number of the blocked
state and its conjugate state to 0.5 and thus re-establishes
artificially the time-reversal symmetry (see, e.g.,
Ref.~\cite{Perez_2008}). In that case, one performs self-consistent
calculations as one would do for the ground-state of an even-even
nucleus.

The SIII parametrization~\cite{Beiner_1975} of the Skyrme effective 
interaction has been chosen since it has been reported to yield very
good nuclear spectroscopic properties in early self-consistent
calculations (see, e.g., Ref.~\cite{Flocard_1973b,Libert_1982}). It has
been shown to meet with a reasonable success in the reproduction of
the spin and parity of odd-$A$ nuclei in the systematic study
of Ref.~\cite{Bonneau_2007}. It is still used in recent studies for
instance in Refs.~\cite{Koh_2016,Niyti_2017,Belabbas_2017,Adel_2017}. 

As already mentioned within our BCS framework, the pairing interaction
is approximated using a spin-singlet seniority force. Its matrix
elements $g^{(q)}$  for the charge state $q$
are given in terms of a parameter $G_q$ and the corresponding number
of particles $N_q$, according to a parametrisation introduced in
Ref.~\cite{Bonche_1985} 
\begin{equation}\label{105}
g^{(q)} = - \frac{G_q}{11 + N_q} \,.
\end{equation}
Indeed, since we are dealing with heavy nuclei not too far from the
valley of stability, we content ourselves
by dealing only with $\lvert T_z \rvert = 1$ pairing. Moreover, we
note that there is a priori no reason for the residual interaction to be such that 
$g^{(n)} = g^{(p)}$
since these matrix elements 
depend on the corresponding different mean fields. Moreover the truncated
single-configuration spaces on which these residual interactions are
projected are different and finally, one must account for the
Coulomb anti-pairing effect (see, e.g., Ref.~\cite{Anguiano_2001}). 

When solving the BCS equations, all single-particle states with
energies up to 6~MeV above the Fermi level are taken into account with a
smoothing factor $\mu = 0.2$~MeV as prescribed in Ref.~\cite{Pillet_2002}.
 
As mentioned earlier, the adiabatic moments of inertia have been
evaluated according to the Inglis-Belyaev formula~\cite{Belyaev_1961} 
\begin{align}\label{106}
\mathcal I = & \sum_{k,l > 0} \dfrac{\vert\langle k\vert\widehat{j}_{+}\vert
  l\rangle\vert^{2}}{(E_{k}+E_{l})}(u_{k}v_{l}-u_{l}v_{k})^{2}
\nonumber \\
& + \frac{1}{2} \sum_{k,l>0}\dfrac{\vert\langle
  k\vert\widehat{j}_{+}\vert\overline{l}\rangle\vert^{2}}{(E_{k}+E_{l})}(u_{k}v_{l}-u_{l}v_{k})^{2}
\end{align}
In this expression, 
the first sum runs on all canonical basis states $k$
such that the projection on the symmetry axis $K_k$
of their total angular momentum is positive
while the sum on the states $l$ is restricted
in practice to states such that $K_l = K_k - 1$. The second sum is
limited to states $k$ and $l$ such that $K_k =
K_l = 1/2$. Furthermore in this
equation $u_m$ and $v_m$ are the absolute values of the BCS probability
amplitudes for the single-particle state $m$ to be empty or filled,
respectively. 

\section{Some aspects of our calculations \label{Technical aspects of the calculations}}

\subsection{Selection of nuclei to be considered}

\begin{figure}[h]
	\includegraphics[angle=0,keepaspectratio=true,scale=0.7]{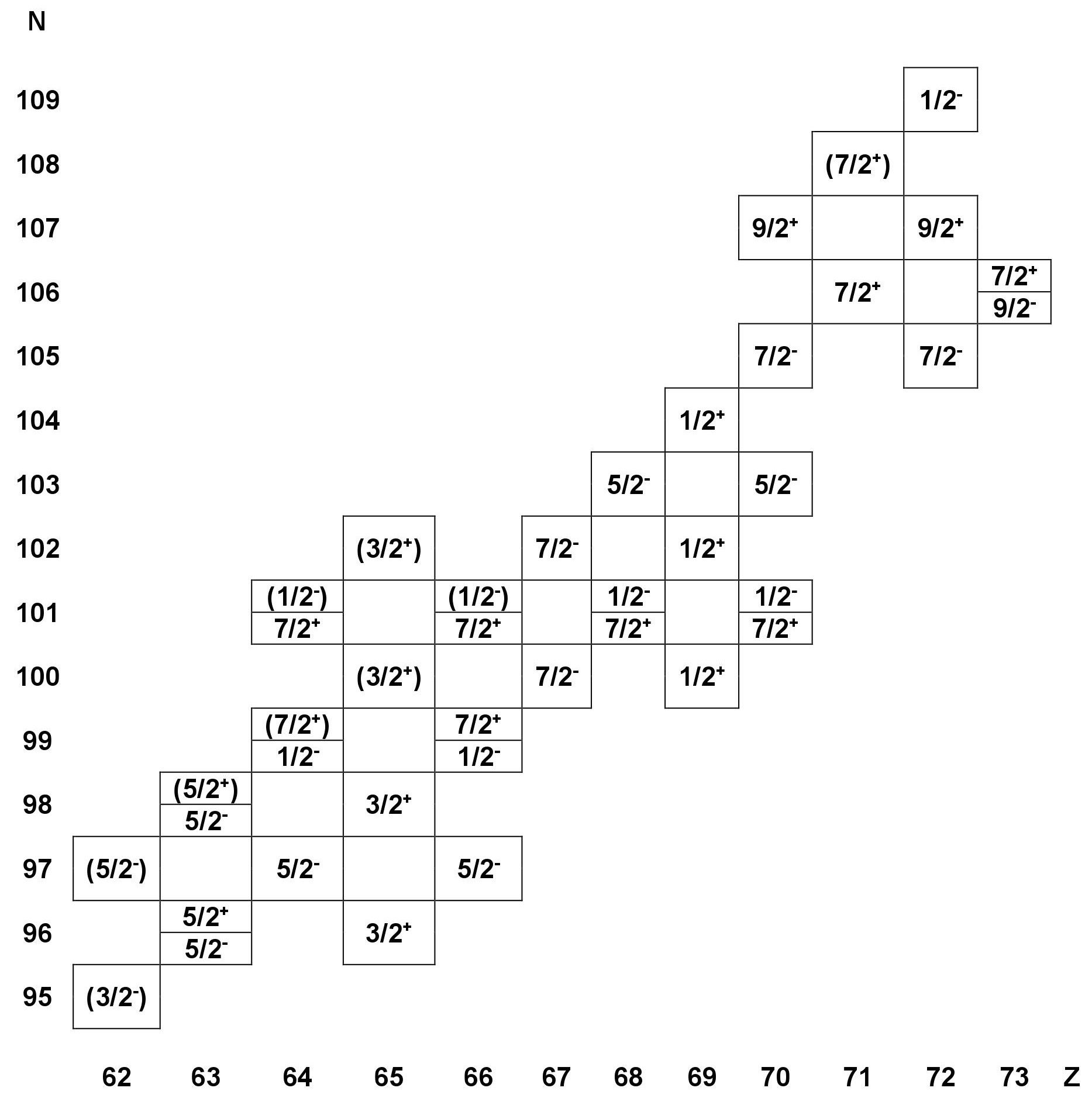}	
	\caption{The nuclear mass region of interest with a total of 24 even-even, 
		17 odd-neutron and 14 odd-proton nuclei considered in this study.
		The ground state experimental quantum 
		numbers  $I^{\pi}$ are displayed (as discussed in the text, they are 
		assumed here to correspond to the $K^{\pi}$ quantum numbers). Whenever 
		our lowest-energy solutions $K^{\pi}$ values are inconsistent with the 
		data, two sets of quantum numbers are displayed. In
                each box the upper panel 
		corresponds to the data, where the use of parenthesis means that these 
		numbers are simply assumed, whereas the lower panel corresponds to our 
		calculated ground state solutions.}
	\label{figure: nuclear region}
\end{figure}

We have included in our study a total of 24 even-even, 17 odd-neutrons
and 14 odd-protons rare-earth nuclei (see Figure~\ref{figure: nuclear region}). Most of the
selected even-even nuclei fulfill the following condition (see Table~\ref{table: nuclear properties even-even}) 
\eq{
\frac{E(4^+)}{E(2^+)} \ge 3.3
}
whereby $E(2^+)$ and $E(4^+)$ are the excitation energies of the first
$2^+$ and $4^+$ states, respectively. This is meant to limit our
sample to well and rigidly deformed nuclei.

\begin{table*}
	\caption{Some nuclear properties for all even-even nuclei considered in this work.
		The second column shows the experimental energy ratio of the first $2^+$ and $4^+$ states 
		while the third and fourth columns 
		show the experimental average between the three-point-mass formulas centered around two neighbouring odd-mass nuclei
		for a given even-even nucleus.
		The calculated charge radii $r_{ch}^{(th)}$ (in fm$^2$) in the fifth column 
		are compared to experimental values $r_{ch}^{(exp)}$ of the next column.
		These experimental data were taken from Ref.~\cite{Fricke_1995} while the numbers in parentheses were taken from
		Ref.~\cite{Angeli_2013}.
		The last two columns are the charge quadrupole moments with experimental values taken from Ref.~\cite{Raman_2001}.
	}.
	\label{table: nuclear properties even-even}
	\begin{ruledtabular}
		\begin{tabular}{*{8}c}
			Nucleus & $E(4^+)/E(2^+)$& $\Delta_{n}^{(3,ave)}$ [MeV]& $\Delta_{p}^{(3,ave)}$ [MeV]& 
			$r_{ch}^{(th)}$ [fm] & $r_{ch}^{(exp)}$ [fm]&	$Q_{20}^{(th)}$ [b] &	 $Q_{20}^{(exp)}$ [b] \\
			\hline
			%%			\cline{3-5}
			%%			&&	Calc.&	Ref.~\cite{Fricke_1995}	& Ref.~\cite{Angeli_2013} &\\
			%%			\hline
			$^{156} \rm Sm$ &	3.290&	0.672&	0.608&	5.144 & -- & 6.903&	--	\\
			$^{158} \rm Sm$ &	3.301&	0.581&	0.558&	5.161 & -- & 7.084&	--	\\
			$^{160} \rm Sm$ &	3.291&	0.623&	0.529&	5.178 & -- & 7.186&	--	\\
			$^{160} \rm Gd$ &	3.300&	0.680&	0.576&	5.191 & 5.174 (5.1734) &	7.276 &	7.265 (42)	\\
			$^{162} \rm Gd$ & 	3.302&	0.658&	0.497&	5.208 & -- & 7.420 &	--	\\
			$^{164} \rm Gd$ &	3.300&	0.706&	0.639&	5.225 & -- & 7.538&	--	\\
			$^{166} \rm Gd$ &	3.297&	0.668&	0.605&	5.240 & -- & 7.605&	--	\\
			$^{162} \rm Dy$ &	3.290&	0.786&	0.642&	5.219 & 5.196 & 7.420&	7.33 (8)	\\
			$^{164} \rm Dy$ &	3.300&	0.679&	0.538&	5.236 & 5.224 & 7.597&	7.503 (33)	\\
			$^{166} \rm Dy$ &	3.310&	0.653&	0.557&	5.253 & -- & 7.724&	--	\\
			$^{168} \rm Dy$ &	3.313&	0.576&	0.576&	5.268 & -- & 7.796&	--	\\
			$^{168} \rm Er$ &	3.309&	0.647&	0.555&	5.281 & 5.272 (5.2644) &	7.890&	7.63 (7)	\\
			$^{170} \rm Er$ &	3.309&	0.622&	0.505&	5.297 & 5.286 (5.2789) &	7.976&	7.65 (7)	\\
			$^{172} \rm Er$ &	3.315&	0.572&	0.505&	5.305 & -- & 7.738&	--	\\
			$^{170} \rm Yb$ &	3.293&	0.749&	0.678&	5.308 & 5.286 (5.2853) &	7.983&	7.63 (9)	\\
			$^{172} \rm Yb$ &	3.305&	0.626&	0.570&	5.323 & 5.301 (5.2995) &	8.067&	7.792 (45)	\\
			$^{174} \rm Yb$ &	3.309&	0.536&	0.527&	5.331 & 5.317 (5.3108) &	7.786&	7.727 (39)	\\
			$^{176} \rm Yb$ & 	3.310&	0.565&	0.485&	5.341 & 5.321 (5.3215) &	7.566&	7.30 (13)	\\
			$^{178} \rm Yb$ &	3.310&	0.607&	0.685&	5.352 & -- & 7.431& --	\\
			$^{176} \rm Hf$ &	3.285&	0.677&	0.686&	5.350 & 5.331 (5.3286) &	7.514&	7.28 (7)	\\
			$^{178} \rm Hf$ &	3.290&	0.635&	0.629&	5.359 & 5.338 (5.3371) &	7.243&	6.961 (43)	\\
			$^{180} \rm Hf$ &	3.307&	0.578&	0.626&	5.371 & 5.349 (5.3470) &	7.094&	6.85 (9)	\\
			$^{182} \rm Hf$ &	3.295&	0.503&	0.555&	5.379 & -- (5.3516) &	6.855&	--	\\
			$^{180} \rm W$ & 	3.260&	0.717&	0.642&	5.379 & -- (5.3491) &	6.965&	6.53 (18)	\\	
		\end{tabular}
	\end{ruledtabular}
\end{table*}

It has been shown that the BCS approach is a bad approximation for low
pairing-correlation regimes (see, e.g., Ref.~\cite{Zheng_1992}). This
is due to the non-conservation of the particle number inherent to the
BCS ansatz. Therefore we chose here to consider odd-$N$ (odd-$Z$ resp.)
nuclei such that their \textit{experimental} pairing gap satisfies 
$\Delta_n> 0.45$~MeV ($\Delta_p>0.45$~MeV resp.). 
These gaps are defined here as the three-point
mass differences centered on a nucleus having an odd number of
neutrons given in Eq.~(\ref{104}) and in a  similar fashion for protons.

In what follows we will need to estimate from the data, pairing gaps
for even-even nuclei. This will be achieved by taking the average of
the $\Delta_n$ ($\Delta_p$ resp.) between the values obtained as above discussed of the two neighbouring odd-$N$
(odd-$Z$ resp.) isotopes (isotones resp.). 

The relevance of such energy differences is contingent upon the
quality of calculated binding energies for each member of the
considered triplet of nuclei, with respect to experimental data. As
shown in the Appendix, whereas our calculated binding energies are
slightly too low in absolute value (by about 4.5~MeV), such a
discrepancy is found to be the same for all nuclei irrespective of the
parity of the nucleon number. This provides a much needed necessary
condition for the estimate of our OES energies.

\subsection{Some calculational details}
The single-particle wave functions of the canonical basis
are expanded on the axially-deformed
harmonic-oscillator basis states. The expansion is truncated following
the prescription of Ref.~\cite{Flocard_1973} in terms of the axial and
perpendicular harmonic-oscillator quantum numbers $n_z$ and $n_{\bot}$
as 
\eq{
\hbar \omega_{\bot} (n_{\bot}+1) + \hbar \omega_z (n_z + \frac{1}{2})
\leq \hbar \omega_0 (N_0 + 2)
}
whereby $\omega_z$ is the
angular frequency in the $z$-direction chosen as the symmetry axis and
$\omega_{\bot}$ is the oscillator frequency in the perpendicular $x-y$
plane, while $\omega_0^3 = \omega_{\bot}^2 \omega_z$
defines the associated spherical oscillator frequency $\omega_0$. In this
study we chose $N_0 = 14$. 

The harmonic-oscillator parameters $b = \sqrt{(m \omega_0)/\hbar}$
(where $m$ is the mean nucleon mass) and  $q = \omega_{\bot}/\omega_z$ are
optimized in order to yield the lowest-energy solution for the
ground-state of the 24 even-even rare-earth nuclei. 
The $b$ and $q$ values for odd-mass nuclei considered in our
calculations are simply the average of the values for the neighbouring even-even
isotopes (isotones resp.) for odd-$N$ (odd-$Z$ resp.) nuclei. 
Numerical integrations are performed using the Gauss-Hermite
quadrature in the $z$-axis and the Gauss-Laguerre quadrature in the
perpendicular plane with 50 and 16 integration points, respectively.

\subsection{Choice of the rare-earth region}

As discussed in Section~II, the relevance of our fits 
is contingent upon the condition of considering rigidly deformed 
nuclei to avoid the bias introduced by quantal shape fluctuations
invalidating both the consideration of a single BCS wavefunction as a
valuable ground-state description and the pollution of first $2^+$ energies
by non-rotational collective modes. On the other hand one should have
at one's disposal an as large as possible sample of nuclei satisfying
this condition.

Two nuclear regions are available a priori: the rare-earth and the
actinide nuclei. The actinide nuclei stable enough to generate
reliable and accurate mass and spectroscopic data is cut-off as well
known by their fission instabilities upon increasing the fissility
parameter. This leaves the single possibility to consider the
rare-earth region. There are 16 even-even isotopes from $Z=62$ to
$Z=72$ which have a ratio of excitation energies of their first $4^+$
and $2^+$ states equal to or larger than 3.3. All these nuclei,
sharing such good rotational properties, have been included in our
sample. They have been complemented by 8 other isoptopes for which
this ratio is close to the 3.3 value.

\section{Results of the fits}

\subsection{Fit based on odd-even mass differences}

To perform this  fit we have computed explicitly the $\Delta_q^{(3)}$
values from the energies of Hartree-Fock plus BCS solutions involving
the three nuclei belonging to the relevant isotopic (or isotonic)
series. These energies are compared with the
experimental ones as given in Ref.~\cite{Chart}. For
an odd-mass nucleus, the lowest-energy solution 
is not necessarily 
obtained by blocking the single-particle state corresponding
to experimental nuclear spin and parity quantum numbers. However, as
seen in Figure~\ref{figure: nuclear region}, in most  
cases (24 out of 31) our calculations yield ground-state spin and parity values
consistent with the data. This confirms the good spectroscopic quality of
the SIII parametrization as already discussed in
Ref.~\cite{Bonneau_2007}. In view of this, we have consistently
considered in our fit the energies of the solutions corresponding to
the experimental $I^{\pi}$ configurations. 

The average rms deviations of $\Delta_n^{(3)}$ and $\Delta_p^{(3)}$ are
displayed in Table~\ref{table: chi OES} on a mesh of relevant
($G_n,G_p$) values. As a first striking result one finds
that the quality of the fit for $\Delta_n^{(3)}$ does not depend significantly on
the values of $G_p$ (and $\Delta_p^{(3)}$ on $G_n$) in the
retained range of parameters $G_n$ 
and $G_p$. In other words, one can perform independent fits of
$\Delta_q^{(3)}$ with respect to $G_q$, provided that one has chosen
a value deemed reasonable for the parameter $G_{\overline{q}}$
associated to the other charge state $\overline{q}$. 

As a result, it appears that the optimum pairing strengths should be in
the vicinity of the $G_n = 16$ MeV and $G_p = 15$ MeV values
for which
$\Delta_n^{(3)}$ is reproduced within 87 keV and
$\Delta_p^{(3)}$ within 182 keV (see Table~\ref{table: chi OES}). 

To yield a specific set of values for ($G_n,G_p$) we have minimized a
$\chi^2$ function combining all odd-$A$ (i.e. odd-$N$ together with
odd-$Z$) calculated results through the expression 
\eq{
\chi^2 = \frac{1}{31} \Big[ \sum_{i=1}^{17} \Big(\Delta_{n,i}^{(th)} -
\Delta_{n,i}^{(exp)} \Big)^2 
+ \sum_{j=1}^{14} \Big(\Delta_{p,j}^{(th)} - \Delta_{p,j}^{(exp)}
\Big)^2 \Big]  
}
where $\Delta_{q,k}^{(th)}$ and $\Delta_{q,k}^{(exp)}$ denote the
calculated and experiment odd-even (three-point) energy differences,
respectively, of the $k^{\rm th}$ nucleus for the charge state $q$.

The corresponding average rms deviations are displayed in
Table~\ref{table: weighted_chi_square_OES}. A polynomial regression of
the third order shows that the minimum is located at $G_n =
16.10$~MeV and $G_p = 14.84$~MeV. 

There is seemingly some arbitrariness in mixing in a single rms
quality indicator, the neutron and proton odd-even mass differences
(with relative weights merely fixed by the numbers of considered
nuclei which happen in our case to be not too different). This does
not turn out to be a problem as demonstrated in the following
way. Taking stock of the already noted independence of the fit of
$G_n$ upon fixing any reasonable value of $G_p$ (and conversely for
the fit of $G_p$ with a reasonable value of $G_n$) we made a
one-dimensional fit of $G_n$ with $G_p = 15$ MeV and  
a one-dimensional fit of $G_p$ with $G_n = 16$ MeV. The resulting
optimal values of $G_n$ (in the first case) and $G_p$ (in the
second case) were found indeed very close to what has been obtained in
the two-dimensional fit. Namely, we found $G_n(G_p= 15) = 16.06$ MeV
and $G_p(G_n= 15) = 15.08$ MeV, which corresponds to the previous
values up to 0.25 \% for neutrons and 1.2 \% for protons.

\begin{table}
	\caption{Average root mean square deviations (in keV) between calculated
		odd-even mass differences $\Delta_n^{(3)}$ and $\Delta_p^{(3)}$
		for different sets of pairing strengths $G_n$ and $G_p$
		(in MeV) } 
	\label{table: chi OES}
	\begin{ruledtabular}
		\begin{tabular}{c*{2}c|*{2}c|*{2}c|*{2}c}
			\multirow{3}{*}{$G_p$} & \multicolumn{8}{c}{$G_n$}\\
			\cline{2-9}
			& \multicolumn{2}{c}{14} & \multicolumn{2}{c}{15} & \multicolumn{2}{c}{16} & \multicolumn{2}{c}{17} \\
			\cline{2-9}
			& $\Delta_n^{(3)}$ & $\Delta_p^{(3)}$ & $\Delta_n^{(3)}$ & $\Delta_p^{(3)}$  & $\Delta_n^{(3)}$ & $\Delta_p^{(3)}$ & 
			$\Delta_n^{(3)}$ & $\Delta_p^{(3)}$  \\
			\hline
			14 & 267 &  288  & 191 & 290 & 83 & 287 & 220 & 286 \\
			\hline
			15 & 282 & 172 & 191 & 182 & 87 & 182 & 224 & 194 \\
			\hline
			16 & 279 & 262 & 189 & 284 & 84 & 289 & 227 & 293 \\
		\end{tabular}
	\end{ruledtabular}
\end{table}

%UPDATED
\begin{table}
	\caption{Same as Table~\ref{table: chi OES} for charge averaged root mean square deviations (in keV) 
		between calculated and experimental odd-even mass differences.}
	\label{table: weighted_chi_square_OES}
	\begin{ruledtabular}
		\begin{tabular}{*{5}c}
			\multirow{2}{*}{$G_p$ }& \multicolumn{4}{c}{$G_n$ } \\
			\cline{2-5}
			& 14 & 15 & 16 & 17 \\
			\hline
			14 & 276.68 & 240.80 & 202.43 & 251.96  \\
			\hline
			15 & 238.68 & 186.99 & 138.24 & 210.98  \\
			\hline
			16 & 271.45 & 236.67 & 203.93 & 258.90  \\
		\end{tabular}
	\end{ruledtabular}
\end{table}

\subsection{Fit based on moments of inertia}

This second fit is performed for all the 24 even-even nuclei
in the rare-earth region which are shown in Figure~\ref{figure: nuclear region}.
As mentioned earlier, the moments of inertia calculated according to
the Inglis-Belyaev formula~\cite{Belyaev_1961} are 
multiplied~\cite{Libert_1999} by a constant $\alpha = 1.32$ to take
into account the so-called Thouless-Valatin corrective terms.  

As well known, because of the 
angular-momentum dependence of the moments of inertia, one has to specify
which definition is retained to evaluate them from data. However the
differences between various reasonable choices are minimal since we
focus here on the first $2^{+}$ state. Here we have defined the moment
of inertia for the rotational-band state of angular momentum 
$I\hbar$ 
from the difference between the incoming and outgoing %outcoming
gamma transition energies corresponding to this state. It is
proportional 
to the inverse of the moment of inertia. We have
thus compared our adiabatic moments of inertia with 
\eq{
\mathcal I^{(\rm exp)} = {4 \hbar^2}/\big[E(4^+) - 2E(2^+)\big]
}
where $E(2^+)$ and $E(4^+)$ are experimental~\cite{Chart} excitation
energies of the first $2^+$ and $4^+$ ground-band states,
respectively.

The average rms deviations between calculated and experimental values are
tabulated in Table~\ref{table: chi square MOI}. Similarly to what has
been obtained with the fit based on odd-even mass differences the best
values in the considered grid are obtained for $G_n = 16$ MeV
and $G_p = 15$ MeV where the rms deviation is found to be 1.75
$\hbar^2 \cdot \rm MeV^{-1}$.

We have obtained the optimal values of $G_n$ and $G_p$ through a cubic
polynomial regression approach to obtain $G_n = 16.27 $ MeV and
$G_p = 15.26$ MeV, which are very close to the values obtained in the previous fit.

\begin{table}
  \caption{Average root mean square deviations of moment of inertia in $\hbar
    ^2 \cdot \rm MeV^{-1}$ unit for
    even-even rare-earth nuclei as a function of pairing strengths.} 
  \label{table: chi square MOI}
  \begin{ruledtabular}
    \begin{tabular}{*{6}c}
      \multirow{2}{*}{$G_p$ [MeV]}& \multicolumn{5}{c}{$G_n$ [MeV]} \\
      \cline{2-6}
      & 14 & 15 & 16 & 17 & 18 \\
      \hline
      13 & 16.34 & 11.35 & 6.17 & 2.28 & 2.97 \\
      \hline
      14 & 13.93 & 8.82 & 3.59 & 1.96 & 5.39 \\
      \hline
      15 & 11.73 & 6.52 & 1.75 & 4.17 & 7.96 \\
      \hline
      16 & 9.86 & 4.25 & 2.36 & 6.39 & 10.26 \\
      \hline
      17 & 8.38 & 3.47 & 3.96 & 8.31 & 12.22 \\
    \end{tabular}
  \end{ruledtabular}
\end{table}

\subsection{Pairing strengths derived from BCS calculations on even-even nuclei}

As already discussed, in many earlier calculations the seniority force parameters 
have been fitted from BCS solutions involving merely even-even nuclei. 
The pairing force intensities have been adjusted so that some calculational results 
were assimilated with odd-even mass differences extracted from experimental nuclear mass tables 
(see, e.g., the analysis of Ref.~\cite{Beiner_1970}). 

In this paper, we want to perform similar fits for the sake of comparison with these approaches.
In our case, these experimental energy differences were obtained for a given even-even nucleus 
by averaging the quantities  $\Delta_q^{(3)}$  
between the two adjacent odd-$N$ nuclei in the isotopic series for the fit of $G_n$ and 
the two adjacent odd-$Z$ nuclei in the isotonic series for the fit of $G_p$. 

We have also mentioned in Section II.A, that two approaches for the fit have been followed. 
In one case, the pairing strengths have been adjusted so as to reproduce the above data 
by some appropriate quasi-particle energies $E_{qp}^{(k)}$ (see Eq.~\ref{102}). 
In the other case one has fitted directly the BCS pairing gaps $\Delta_q$. 

To be consistent with what has been done in Section V.A
we have retained the quasi-particle states having the lowest quasi-particle energy 
for the quantum numbers $K^{\pi}$ corresponding to the experimental ground state configuration $I^{\pi}$. 

As a result we expect for reasons previously discussed, to obtain fitted pairing strength parameters 
smaller than what was obtained by explicit calculations of $\Delta_q^{(3)}$ quantities. 
The aim of this Section is to estimate to which extent they are underestimated. 

In the care where quasi-particle energies are used in the fit, we have obtained the results displayed in 
Table~\ref{table: quasiparticle fit} for the rms energy differences between calculated and experimental $\Delta_q^{(3)}$ energies. Table~\ref{table: weighted_chi_square_Eqp} displays the results of a combined (proton and neutron)
$\chi^2$ analysis similar to what has been done in Section V.A. 
It yields optimal values $G_n = 14.78$ MeV and $G_p = 12.36$ MeV. 
The neutron strength $G_n$ is indeed found moderately lower than the
one obtained from exact $\Delta_q^{(3)}$ calculations, while it is 
largely quenched for protons.

It is to be noted that, while this set of optimal pairing strengths
yields a remarkable agreement for odd-neutron gaps as seen in
Table~\ref{table: quasiparticle fit}, it is nevertheless inconsistent
with the fit based on moment of inertia. 

The same type of analysis has been made when the fit is performed on pairing-gap values. 
The rms deviations obtained for the OES differences are displayed on Table~\ref{table: delta BCS fit} 
while the results of the combined $\chi^2$ analysis are displayed on Table~\ref{table: weighted_chi_square_deltaq}. 
We obtain the following set of seniority strength parameters: $G_n = 15.40$ MeV and $G_p = 13.67$ MeV.  
The expected quenching effect on the $G_q$ values is present 
but less important than what was observed when fitting on the quasiparticle energies. 
This can be understood since we omit in the former case the contribution of the $(e_k - \lambda)^2$ term present in the latter.	%, 
%thus minimizing the too high point of comparison with  $\Delta_q^{(3)}$ data.

To quantify in a concrete example the consequence of the approximation made by 
determining pairing strengths from such calculations on even-even nuclei, 
we have computed the moments of inertia for our sample of 24 even-even nuclei 
with the seniority-force parameters obtained in the quasiparticle-energy version of our fit. 
The results are displayed on Table~\ref{table: effect of pairing on MOI}. 
When applying as we should the Thouless-Valatin correction to the Inglis-Belyaev results 
we found as expected a huge overestimation of the moments of inertia. 
It is a remarkable coincidence that without this necessary correction 
the results are found in a very good agreement with the data. 
That could have possibly prevented authors who discarded this
correction and made a pairing-strength fit merely on odd-even mass
differences out of even-even nuclear solutions from
realizing that they were artificially lowering the strength of their pairing residual 
interaction. This should of course yield important consequences on a 
further description of other properties  affected significantly by the 
level of pairing correlations.

\begin{table}
	\caption{Average root mean square deviations between calculated and experimental odd-even mass differences 
		for different sets of pairing strengths based on quasi-particle energies.}
	\label{table: quasiparticle fit}
	\begin{ruledtabular}
		\begin{tabular}{c*{2}c|*{2}c|*{2}c|*{2}c}
			\multirow{3}{*}{$G_p$} & \multicolumn{8}{c}{$G_n$}\\
			\cline{2-9}
			& \multicolumn{2}{c}{13}& \multicolumn{2}{c}{14} & \multicolumn{2}{c}{15} & \multicolumn{2}{c}{16} \\	
			\cline{2-9}
			& $\Delta_n^{(3)}$ & $\Delta_p^{(3)}$ & $\Delta_n^{(3)}$ & $\Delta_p^{(3)}$  & $\Delta_n^{(3)}$ & $\Delta_p^{(3)}$ & 
			$\Delta_n^{(3)}$ & $\Delta_p^{(3)}$  \\
			\hline
			11 & 210.60& 174.89& 137.79 &  177.48 &  82.86&  177.64&  203.37&	179.15	\\	
			\hline
			12 & 208.19& 105.15& 138.36 &  104.76 &  82.57&  106.41&  200.26&  106.58	\\
			\hline
			13 & 211.99& 215.88& 138.62 & 78.32 & 82.83 & 73.27 & 201.45 & 75.84 \\	
			\hline
			14 & 212.7 & 215.88 & 138.80 & 216.90 & 80.10 & 214.60 & 204.70 & 213.10 \\
			\hline
			15 & 177.67 & 404.53 & 96.10 & 412.0 & 113.20 & 403.00 & 216.70 & 410.70 \\	
		\end{tabular}
	\end{ruledtabular}
\end{table}

\begin{table}
	\caption{Average root mean square deviations (in keV) based on a fit using quasi-particle energy, $E_{qp}^{(k)}$.}
	\label{table: weighted_chi_square_Eqp}
	\begin{ruledtabular}
		\begin{tabular}{*{5}c}
			\multirow{2}{*}{$G_p$ [MeV]}& \multicolumn{4}{c}{$G_n$ [MeV]} \\
			\cline{2-5}
			& 13 & 14 & 15 & 16 \\
			\hline
			11 & 193.57& 158.88 & 138.60& 191.64	\\
			\hline
			12 & 164.93& 122.71 & 95.24 & 160.41	\\
			\hline
			13 & 160.47 & 112.58 & 78.20 & 148.87	\\
			\hline
			14 & 214.31 & 182.09 & 162.00 & 208.90	\\
			\hline
			15 & 312.42 & 286.44 & 296.00 & 314.30	\\
			%			\hline
		\end{tabular}
	\end{ruledtabular}
\end{table}

\begin{table}
	\caption{Average root mean square deviations (in keV) between calculated and experimental odd-even mass differences 
		for different sets of pairing strengths based on $\Delta_{\mbox{BCS}}$.}
	\label{table: delta BCS fit}
	\begin{ruledtabular}
		\begin{tabular}{c*{2}c|*{2}c|*{2}c|*{2}c}
			\multirow{3}{*}{$G_p$} & \multicolumn{8}{c}{$G_n$}\\
			\cline{2-9}
			& \multicolumn{2}{c}{14}& \multicolumn{2}{c}{15} & \multicolumn{2}{c}{16} & \multicolumn{2}{c}{17} \\	
			\cline{2-9}
			& $\Delta_n^{(3)}$ & $\Delta_p^{(3)}$ & $\Delta_n^{(3)}$ & $\Delta_p^{(3)}$  & $\Delta_n^{(3)}$ & $\Delta_p^{(3)}$ & 
			$\Delta_n^{(3)}$ & $\Delta_p^{(3)}$  \\
			\hline
			11 & 367.39& 462.82& 167.85 &  466.14 &  131.21&  472.59&  328.78&	482.57	\\	
			\hline
			12 & 369.11& 325.88& 171.39 &  329.11 &  130.14&  332.58&  328.39&  341.01	\\
			\hline
			13 & 372.21& 191.39& 168.59 & 192.80 & 132.64 & 193.68 & 329.12 & 194.79 \\	
			\hline
			14 & 372.67 & 346.72 & 160.16 & 345.03 & 135.99 & 343.02 & 337.51 & 340.79 \\
			\hline
			15 & 372.36 & 346.72 & 160.16 & 345.03 & 135.99 & 343.02 & 337.51 & 340.79 \\	
		\end{tabular}
	\end{ruledtabular}
\end{table}

\begin{table}
	\caption{Average root mean square deviations (in keV) between calculated and experimental pairing gap based on a fit to $\Delta_{\mbox{BCS}}$.}
	\label{table: weighted_chi_square_deltaq}
	\begin{ruledtabular}
		\begin{tabular}{*{5}c}
			\multirow{2}{*}{$G_p$ [MeV]}& \multicolumn{4}{c}{$G_n$ [MeV]} \\
			\cline{2-5}
			& 14 & 15 & 16 & 17 \\
			\hline
			11 & 417.84 & 350.33 & 346.81& 412.90	\\
			\hline
			12 & 348.17 & 262.38 & 252.53 & 334.76 \\
			\hline
			13 & 295.95 & 181.10 & 165.99 & 270.43 \\
			\hline
			14 & 285.20 & 155.73 & 137.40 & 256.73 \\
			\hline
			15 & 359.77 & 268.98 & 260.92 & 339.16 \\
			%			\hline
		\end{tabular}
	\end{ruledtabular}
\end{table}

\begin{table}
	\caption{Moment of inertia (in units of $\hbar^2/\mbox{MeV}$)
          calculated using the Inglis-Belyaev formula with 
          Thouless-Valatin correction $\gimel_{TV}$ using two sets of ($G_n$,$G_p$) pairing
          strengths (in MeV): $(16,15)$ from a fit to OES,
          $(14.8,12.4)$ from a fit to quasiparticle energies.
      		Experimental moments $\gimel_{exp}$ are also given. In the (14.8,12.4) case we have added 
      		the uncorrected Inglis-Belyaev values $\gimel_{IB}$ for the sake of 
      		comparison with $\gimel_{exp}$ as discussed in the text. Finally in the 
      		(16,15) case, we have displayed the Thouless-Valatin corrected values 
      		$\gimel_{TV}^{exact}$ obtained when treating exactly the Coulomb exchange contribution.}
	\label{table: effect of pairing on MOI}
	\begin{ruledtabular}
		\begin{tabular}{*{6}c}
			\multirow{2}{*}{Nucleus}& 	\multicolumn{2}{c}{(16,15)}&	
					\multicolumn{2}{c}{(14.8,12.4)}&	\multirow{2}{*}{$\gimel_{exp}$}	\\
					\cline{2-3}	\cline{4-5}
				&	$\gimel_{TV}$&	$\gimel_{TV}^{exact}$&	$\gimel_{IB}$& 	$\gimel_{TV}$&	\\
			\hline
			$^{156}$Sm&	41.21&	42.40&		39.30&	51.88&	40.846	\\
			$^{158}$Sm&	41.54&	42.70&		39.91&	52.68&	42.239	\\
			$^{160}$Sm&	44.49&	45.62&		44.58&	58.84&	43.716	\\
			$^{160}$Gd&	39.69&	41.40&		39.96&	52.75&	40.816	\\
			$^{162}$Gd&	44.58&	46.57&		47.29&	62.42&	42.918	\\
			$^{164}$Gd&	42.95&	45.13&		42.47&	56.06&	41.973	\\
			$^{166}$Gd&	45.79&	48.40&		50.26&	66.35&	44.053	\\
			$^{162}$Dy&	38.05&	39.46&		38.71&	51.10&	38.335	\\
			$^{164}$Dy&	43.34&	44.69&		46.50&	61.38&	41.908	\\
			$^{166}$Dy&	41.32&	42.68&		40.28&	53.17&	39.859	\\
			$^{168}$Dy&	44.00&	45.49&		45.66&	60.28&	40.646	\\
			$^{168}$Er&	39.23&	39.92&		36.58&	48.28&	38.285	\\
			$^{170}$Er&	42.37&	43.33&		43.37&	57.25&	38.854	\\
			$^{172}$Er&	36.72&	37.57&		34.89&	46.06&	39.526	\\
			$^{170}$Yb&	38.64&	39.35&		36.90&	48.71&	36.724	\\
			$^{172}$Yb&	41.35&	42.49&		42.63&	56.27&	38.917	\\
			$^{174}$Yb&	37.13&	38.97&		37.85&	49.96&	39.930	\\
			$^{176}$Yb&	35.73&	37.88&		35.78&	47.22&	37.182	\\
			$^{178}$Yb&	37.80&	40.38&		37.94&	50.09&	36.364	\\
			$^{176}$Hf&	33.87&	34.56&		33.70&	44.49&	35.248	\\
			$^{178}$Hf&	33.46&	34.51&		33.65&	44.42&	33.262	\\
			$^{180}$Hf&	35.22&	36.26&		35.39&	46.71&	32.806	\\
			$^{182}$Hf&	32.06&	33.07&		30.55&	40.32&	31.598	\\
			$^{180}$W&	30.55&	30.69&		29.07&	38.38&	30.666	\\
		\end{tabular}
	\end{ruledtabular}
\end{table}

\begin{table}
	\caption{Optimum pairing strengths (in MeV) obtained from various fitting procedures.}
	\label{table: tabulated pairing strengths}
	\begin{ruledtabular}
		\begin{tabular}{*{1}l*{2}c}
			Fit procedures& 	$G_n$& 	$G_p$	\\
			\hline
			Moment of inertia&	16.27&	15.26	\\
			OES using SCB&		16.10&	14.84	\\
			OES using $\Delta_{\mbox{BCS}}$&	15.40&	13.67	\\
			OES using $E_{qp}$&	14.78&	12.36	\\
		\end{tabular}
	\end{ruledtabular}
\end{table}

\subsection{Comparison with similar attempts to fit the pairing
  residual interaction} 

It is worth comparing our results with those obtained within the OES
protocol in Refs.~\cite{Bertsch_2009,Kortelainen_2012}. In
both , one uses a zero-range density-dependent residual interaction to
define the pairing part of the Energy Density Functional (EDF). For
the particle-hole part in their EDF, the authors of
Ref.~\cite{Bertsch_2009} use the SLy4 parametrization of the Skyrme
interaction \cite{Chabanat_1998} while those of
Ref.~\cite{Kortelainen_2012} start from a previous EDF
parametrization, called UNEDF0~\cite{Kortelainen_2010} to improve it
as a UNEDF1 version. 

Our comparison will be based on the r.m.s. error (in keV) obtained for
neutrons and protons for the three-point energy differences
$\Delta_q^{(3)}$. In Ref.~\cite{Bertsch_2009}, these values are at
best, i.e. within the favored Hartree-Fock-Bogoliubov (HFB) plus
Lipkin-Nogami approach, about 250 keV for both charge states. The
corresponding results in Ref.~\cite{Kortelainen_2012} are 342 keV
(350 keV resp.) for neutrons in $A \geqslant 80$ nuclei with the
UNEDF0 (UNEDF1 resp.) while the corresponding figures are 229 keV
(resp.~248 keV). In our approach now, for $G_n = 16$ MeV and $G_p
= 15$ MeV, we have obtained 87 keV for neutrons and 182 keV for
protons which corresponds to a significant improvement.

Three remarks are in order here. First, the numbers of nuclei included
in the sample of both approaches in Refs.~\cite{Bertsch_2009} and
\cite{Kortelainen_2012} is considerably larger. This does not
constitute necessarily a decisive advantage since one should be a
priori rather selective in any fitting process. Second on Fig. 7 of
Ref. \cite{Bertsch_2009} a significant deformation dependence of the
r.m.s. error for $\Delta_n^{(3)}$ is exhibited. Within the HFB
approach (slightly less good than their HFB plus Lipkin Nogami
approach) the authors of this paper found that the corresponding
r.m.s. error was reduced from 270 keV to 250 keV upon limiting their
sampling to nuclei in our region of interest, namely for nuclei whose
quadrupole deformation parameter $\beta$ was found in the 0.2-0.3
range. Finally, in the section VI of Ref.~\cite{Bertsch_2009}, a
suggestive remark has been made about the the intensity of the proton
residual interaction.These authors found it larger by about $10\%$
than what is obtained for neutrons. The authors rightfully express
that ``the Coulomb interaction in the pairing channel [...] would be
expected to decrease the … strength, not to increase it''. It is to be
noted that we found the reverse effect ($G_n$ significantly larger
than $G_p$) which seems more easily understood. 

\section{Conclusion}

In this paper we have substantiated the statement made in the seminal paper of Bohr, Mottelson and Pines \cite{Bohr_1958} 
that pairing properties could be very well be assessed by correctly reproducing 
both the odd-even energy staggering 
and the moments of inertia of the first members of ground-state rotational band in well-deformed nuclei.
As summarized in Table~\ref{table: tabulated pairing strengths} we found, indeed, 
an excellent agreement between the outputs of the two independent approaches. 

Obtaining this we have also demonstrated that our crude theoretical approach of both properties 
(limitation to seniority force BCS calculations, 
global renormalization of moments of inertia due to the 
Thouless-Valatin corrections as proposed in Ref.~\cite{Libert_1999}, 
simple parametrization of the particle number dependence of the seniority force strength, for instance) 
was most probably accurate enough to describe the properties under study.

We have also shown (see Table~\ref{table: tabulated pairing strengths}) 
that widely used fitting protocols of pairing properties from odd-even energy differences 
deduced merely from solutions for even-even nuclei were by far not appropriate.

Since it is clear that it is simpler to compute moments of inertia  
in even-even nuclei than to compute explicitly odd-even mass differences, 
our results could have a real practical impact on the fit of residual interactions.

There are clearly many points that could be improved, among which the
use of a seniority force and the particle-number breaking character of
the BCS approximation. Both issues are currently tackled within the
so-called Highly Truncated Diagonalization Approach (HTDA) of
Ref.~\cite{Pillet_2002} where a zero-range delta residual interaction
is used within a variational approach on good particle-number trial
wavefunctions. 

One should thus consider that the main physical motivation of this
study is to substantiate the point of principle suggested in
Ref.~\cite{Bohr_1958} about the relevance of OES energies and moments
of inertia to determine the amount of pairing correlations. This point
being made we intend to move forward and perform a fit of more
sophisticated residual interactions to be used within the HTDA
formalism to study spectroscopic properties where an accurate
treatment of pairing plays an important role. This is in particular
the case when studying high-$K$ isomers where the Pauli
blocking effect quenches the pairing correlations in a low regime
where the HF+BCS (or HFB for this matter) approximation is known to be
unsatisfactory (see e.g., Ref~\cite{Zheng_1992}).

Another deficiency is to be quoted. It has been consistently found
here that proton properties were leading to slightly less satisfactory
properties than neutron ones. This can be seen in the rms values in
various fits or similarly the significantly larger--yet small in
absolute terms--differences between the two fits of Section V.A and V.B. 
This might result from the systematic effect on level density around the
Fermi energy of the approximate Slater treatment of the Coulomb
exchange term (see e.g. Ref.\cite{Koh_2017}). Indeed, the approximate
spectra are significantly more compressed than exact ones. 
This yield in the latter case, upon using the same
residual interaction, slightly larger moments of inertia as quantified
in the comparison of Table~\ref{table: effect of pairing on MOI}. Of
course to each Energy Density Functional should correspond a specific
fit of the residual interaction and the exact Coulomb exchange
calculations have been performed here merely for the sake of
illustration of the limit of the EDF in use. It is clear that the
numerical results of our present fit are to be used with a SIII Skyrme
EDF with Coulomb exchange terms in the Slater approximation.

Having pointed out the various limitations of our current approach, 
we think it possible nevertheless to conclude that the remarkable agreement 
between the results of the two fits based on very different physical properties 
should very likely survive at least qualitatively when attempting similar calculations in most advanced theoretical frameworks.

\appendix*
\section{Comparison of calculated and experimental ground-state binding energies}

The binding energy calculated using the Skyrme SIII parametrization
for ground-states of both even-even and odd-mass nuclei are tabulated
in Table~\ref{table: binding energy} and compared to experimental
data~\cite{Wang_2017}. The r.m.s deviations for 24 even-even, 17
odd-neutron and 14 odd-proton nuclei are 4.64 MeV, 4.48 MeV and 4.45
MeV, respectively. This leads to a r.m.s deviation of 4.54 MeV for all
the considered 55 rare-earth nuclei.

One notes therefore a systematic underbinding of our solutions (in
absolute value). This leaves some room for corrections of
various origins, such as truncation of the basis or zero-point
motions. Yet this error is found to be very similar irrespective of
the parity of the neutron and proton numbers. This consistency is very
a important point, in our case, since the OES energies imply
differences between even-even and odd-$N$ even-$Z$ or even-$N$ odd-$Z$
nuclei.

\begin{table*}[h]
\caption{Binding energies (in MeV) calculated using the Skyrme SIII,
  $B_{\rm th}$, and compared to the experimental values $B_{\rm exp}$ from \cite{Wang_2017}.
The ground-state spin and parity quantum numbers of odd-mass nuclei are given in parentheses.}
\label{table: binding energy}
\begin{tabular}{*{3}c|*{3}c|*{3}c}
\hline \hline
\multicolumn{3}{c|}{Even-even nuclei} & \multicolumn{3}{c|}{Odd-$N$
  nuclei} & \multicolumn{3}{c}{Odd-$Z$ nuclei} \\
\hline
Nucleus & $B_{\rm th}$ & $B_{\rm exp}$ &
Nucleus & $B_{\rm th}$ & $B_{\rm exp}$ &
Nucleus & $B_{\rm th}$ & $B_{\rm exp}$ \\
\hline
$^{156}$Sm & 1275.54 & 1279.98 & $^{157}$Sm (3/2$^{-}$) & 1281.18 & 1285.37 & $^{159}$Eu (5/2$^{+}$) & 1295.66 & 1300.09	\\
$^{158}$Sm & 1287.65 & 1292.01 & $^{159}$Sm (5/2$^{-}$) & 1292.97 & 1297.04 & $^{161}$Eu (5/2$^{+}$) & 1307.75 & 1311.99	\\
$^{160}$Sm & 1299.07 & 1303.14 & $^{161}$Gd (5/2$^{-}$) & 1310.52 & 1314.92 & $^{161}$Tb (3/2$^{+}$) & 1311.51 & 1316.09	\\
$^{160}$Gd & 1304.58 & 1309.28 & $^{163}$Gd (7/2$^{+}$) & 1322.77 & 1326.87 & $^{163}$Tb (3/2$^{+}$) & 1324.97 & 1329.37	\\
$^{162}$Gd & 1317.36 & 1321.76 & $^{165}$Gd (1/2$^{-}$) & 1334.21 & 1338.15 & $^{165}$Tb (3/2$^{+}$) & 1337.47 & 1341.45	\\
$^{164}$Gd & 1329.20 & 1333.32 & $^{163}$Dy (5/2$^{-}$) & 1325.49 & 1330.37 & $^{167}$Tb (3/2$^{+}$) & 1349.12 & 1353.03	\\
$^{166}$Gd & 1340.21 & 1344.27 & $^{165}$Dy (7/2$^{+}$) & 1339.16 & 1343.74 & $^{167}$Ho (7/2$^{-}$) & 1353.17 & 1357.77	\\
$^{162}$Dy & 1319.00 & 1324.10 & $^{167}$Dy (1/2$^{-}$) & 1351.87 & 1356.21 & $^{169}$Ho (7/2$^{-}$) & 1366.13 & 1370.43	\\
$^{164}$Dy & 1333.09 & 1338.03 & $^{169}$Er (1/2$^{-}$) & 1367.05 & 1371.78 & $^{169}$Tm (1/2$^{+}$) & 1366.60 & 1371.35	\\
$^{166}$Dy & 1346.22 & 1350.79 & $^{171}$Er (5/2$^{-}$) & 1380.16 & 1384.71 & $^{171}$Tm (1/2$^{+}$) & 1380.75 & 1385.42	\\
$^{168}$Dy & 1358.51 & 1362.90 & $^{171}$Yb (1/2$^{-}$) & 1379.92 & 1384.74 & $^{173}$Tm (1/2$^{+}$) & 1393.87 & 1398.61	\\
$^{168}$Er & 1360.79 & 1365.77 & $^{173}$Yb (5/2$^{-}$) & 1394.23 & 1399.12 & $^{177}$Lu (7/2$^{+}$) & 1421.05 & 1425.46	\\
$^{170}$Er & 1374.32 & 1379.03 & $^{175}$Yb (7/2$^{-}$) & 1407.70 & 1412.41 & $^{179}$Lu (7/2$^{+}$) & 1434.24 & 1438.28	\\	
$^{172}$Er & 1386.78 & 1391.55 & $^{177}$Yb (9/2$^{+}$) & 1420.51 & 1424.85 & $^{179}$Ta (7/2$^{+}$) & 1432.95 & 1438.01	\\
$^{170}$Yb & 1373.10 & 1378.12 & $^{177}$Hf (7/2$^{-}$) & 1420.41 & 1425.17 & \\
$^{172}$Yb & 1387.77 & 1392.76 & $^{179}$Hf (9/2$^{+}$) & 1434.68 & 1438.90 & \\
$^{174}$Yb & 1401.52 & 1406.59 & $^{181}$Hf (1/2$^{-}$) & 1447.45 & 1451.98 & \\
$^{176}$Yb & 1414.66 & 1419.28 & &&&&&\\
$^{178}$Yb & 1427.04 & 1431.63 & &&&&&\\
$^{176}$Hf & 1413.93 & 1418.80 & &&&&&\\
$^{178}$Hf & 1428.29 & 1432.80 & &&&&&\\
$^{180}$Hf & 1441.89 & 1446.29 & &&&&&\\
$^{182}$Hf & 1454.25 & 1458.70 & &&&&&\\
$^{180}$W  & 1439.69 & 1444.58 & &&&&&\\
\hline \hline
\end{tabular}
\end{table*}

\begin{acknowledgments}
N.M.N. and N.A.R. would like to thank Universiti Teknologi Malaysia for financial assistance
through the Potential Academic Staff (grant number Q.J130000.2726.02K70) and 
the Research University Grant (grant number Q.J130000.2626.15J74).
Both also gratefully acknowledge the hospitality extended to them during their visit to the CENBG.
P.Q. warmly thanks UTM for fruitful visits to the Faculty of Science.  
%In this respect the support of the SCAC of the French Embassy in Kuala Lumpur is also to be acknowledged.
Finally, the important support of the SCAC of the French Embassy in Kuala Lumpur is also to be strongly acknowledged.

\end{acknowledgments}

\end{document}